\documentclass[12pt]{article}
\begin{document}


\def\tr{{\hbox{\rm Tr}}}
\def\ex{{\hbox{\rm e}}}
\def\k{{\cal K}}

\newcommand{\beq}{\begin{equation}}
\newcommand{\eeq}{\end{equation}}
\newcommand{\bear}{\begin{eqnarray}}
\newcommand{\eear}{\end{eqnarray}}

\newcommand{\CS}{{\scriptstyle {\rm CS}}}
\newcommand{\CSs}{{\scriptscriptstyle {\rm CS}}}
\newcommand{\ie}{{\it i.e.}}

\newdimen\tableauside\tableauside=1.0ex
\newdimen\tableaurule\tableaurule=0.4pt
\newdimen\tableaustep
\def\phantomhrule#1{\hbox{\vbox to0pt{\hrule height\tableaurule width#1\vss}}}
\def\phantomvrule#1{\vbox{\hbox to0pt{\vrule width\tableaurule height#1\hss}}}
\def\sqr{\vbox{%
  \phantomhrule\tableaustep
  \hbox{\phantomvrule\tableaustep\kern\tableaustep\phantomvrule\tableaustep}%
  \hbox{\vbox{\phantomhrule\tableauside}\kern-\tableaurule}}}
\def\squares#1{\hbox{\count0=#1\noindent\loop\sqr
  \advance\count0 by-1 \ifnum\count0>0\repeat}}
\def\tableau#1{\vcenter{\offinterlineskip
  \tableaustep=\tableauside\advance\tableaustep by-\tableaurule
  \kern\normallineskip\hbox
    {\kern\normallineskip\vbox
      {\gettableau#1 0 }%
     \kern\normallineskip\kern\tableaurule}%
  \kern\normallineskip\kern\tableaurule}}
\def\gettableau#1 {\ifnum#1=0\let\next=\null\else
  \squares{#1}\let\next=\gettableau\fi\next}

\tableauside=1.0ex
\tableaurule=0.4pt

%
%

\thispagestyle{empty}

\begin{flushright}
USC-FT-9-01 \\
hep-th/0107079 
\end{flushright}

\medskip

\begin{center}

{\Large{\bf{Math and Physics\footnotemark}}} 

\medskip\medskip\medskip

{\sc Jos\'e M. F. Labastida}   
\vskip0.5cm

\footnotetext{Talk delivered at the special seminar in honor of F. J.
Yndur\'ain held at Sitges, Spain, on February 9, 2001}

{\it Departamento de F\'\i sica de Part\'\i
culas \\Universidade de Santiago de Compostela\\
15706 Santiago de Compostela, Spain\\
{\rm e-mail: labasti@fpaxp1.usc.es}}         
\end{center}
\bigskip

%
%

\begin{abstract}
I present a brief review on some of the recent developments in
topological quantum field theory. These include topological
string theory, topological Yang-Mills theory and Chern-Simons
gauge theory. It is emphasized how the application of different
field and string theory methods has led to important
progress, opening entirely new points of view in the
context of Gromov-Witten invariants, Donaldson invariants, and
quantum-group invariants for knots and links.

\end{abstract}

%
%

\section{Introduction}

The last two decades constituted a very fruitful time for theoretical
physics. During this period the use of geometrical and topological
methods have been particularly intense, leading to a new type of relation
between physics and mathematics. Beginning in the eighties, we have
witnessed how the most advanced developments in theoretical physics have
led to new results in mathematics. A new type of relation emerged which is
unprecedented in history. Mathematics and physics enjoyed a happy
marriage during hundreds of years and evolved together up to the end of
the XIX$^{\rm th}$ century. Official divorce took place at the beginning
of the XX$^{\rm th}$ century when abstraction started to play a
fundamental role in mathematics. During their happy years new fields in
mathematics were often created because they were needed by physics. In the
twentieth century, however, mathematics was always ready for physics
needs. This was the case, for example, of general relativity and quantum
mechanics. One of the members of the couple did not need the other
anymore and their relation cooled down.

 The eighties constituted a period of reconciliation. Physics had the
courage to tell its old partner were to look at and this did it quite
successfully. Quantum field theory and string theory
started to generate new mathematics, establishing a new type of relation
which hopefully will last for many years. The most remarkable aspect
of this new liaison is that the emerging new mathematics were not produced
because they were needed by physics but because they could lead to
important breakthroughs in mathematics. Topology was the field of
mathematics that was particularly involved in these developments. On the
physics side, the quantum aspects of field and string theory were the
responsible for the establishment of the new connections in mathematics.
All these developments have led to the study of new types of theories now
known as topological field theories and topological string theories.

The motivation for the study of these theories is not only mathematical.
There are reasons that make them also interesting from a physics point of view.
First, they are simpler than theories related to the real world. This
allows to solve them exactly in some cases, generating an important knowledge
on the structure of field and string theory. Second, these theories could be
used as starting points to solve more realistic theories. Third, they
constitute an excellent framework to test physics arguments which are not
firmly supported from a mathematical point of view.

The third motivation is indeed related to the cyclic nature of the new
relation between physics and mathematics. Results in field and string theory
which use arguments based on functional integrals (arguments of
quantum origin) can not be considered in general as firm arguments from a
mathematical perspective. Those results have to be considered as
predictions in mathematics made by physics. Once the predictions are
proposed, mathematicians have to provide the corresponding rigorous
proofs. In a certain sense, mathematicians play the role of experimental
physicists when theoretical physicists make predictions that have to be
tested in the laboratories. Needles to say that the predictions made by
topological field and string theory are much cheaper to test than the
ones made by ordinary theories.

The analogy with experimental physics can be pursued further. Often, from
the test of theoretical predictions in laboratories, new phenomena is
observed that leads to new theoretical studies. Similarly, from the rigorous
proofs of the predictions made from topological field theory and string
theory new facts are encountered that, in turn, lead to further study in
physics. These, eventually, could lead to additional new results in
mathematics, establishing a cyclic structure as in the case of experimental
physics.

A particularly good example to illustrate this cyclic behavior 
is Morse-Witten theory. In 1982 Witten studied supersymmetric quantum
mechanics and supersymmetric sigma models making a prediction on the
existence of an interesting generalization of Morse theory \cite{morse}. This
prediction was later studied by A. Floer, establishing its fundamentals from a
mathematical perspective \cite{floer}. The theory was later extended by
Floer himself to other contexts, motivating further studies by Witten
which, after some pressure by M. Atiyah \cite{atiyah}, culminated in the
formulation of Donaldson theory \cite{donald} in a field theory framework
in 1988
\cite{tqft}.

The developments of the type discussed above have been taking place in field
theories defined in 2, 3 and 4 dimensions, and in the context of string
theory. In this article I will present a brief review of these developments
emphasizing on the recent ones in the context of knot theory. These are
particularly important because they show how useful field and string
theory are to unravel the many faces of the theory of knot and link
invariants. The formalism leads to relations which otherwise would have
been rather hard to discover.

The cases that will be considered in this paper have two common features. In all
of them, first, a particular problem in topology is formulated in terms of
field or string theory. In particular, we will be dealing with Gromov-Witten
invariants, Donaldson invariants and quantum-group invariants. For all these
invariants, topological field theory or topological string theory provide a
representation  in terms of vacuum expectation values of some
``physical observables". Second, field and string theory methods are used to
express these vacuum expectation values in alternative forms, leading to
expressions that involve new integer invariants. In the case of Donaldson
invariants these new integer invariants are the celebrated Seiberg-Witten
invariants. Similarly, in the case of Gromov-Witten invariants one finds that
these can be resumed and expressed in terms of simpler integer invariants. 

For
quantum-group invariants progress evolved in two steps. First, the
quantum-group polynomial invariants, whose coefficients are integer numbers, are
expressed, after a string theory reformulation of Chern-Simons gauge theory, in
terms of new non-integer invariants which are a particular analog of
Gromov-Witten invariants for Riemann surfaces with boundaries. In the second
step these non-integer coefficients are resumed in terms of new 
integer invariants which are related by a simple reformulation to the integer
coefficients of quantum-group invariants. These new invariants can be
interpreted in terms of enumerative geometry and therefore the formalism
assigns a geometrical meaning to the integer coefficients of the quantum-group
polynomial invariants, a question that remained open since their formulation.

The organization of the paper is as follows. In section 2, I briefly define
quantum field theory and I present a brief review of these theories in 2 and
4 dimensions, dealing with Gromov-Witten invariants, and Donaldson and
Seiberg-Witten invariants, respectively. In section 3, I review Chern-Simons
gauge theory and its role in the theory of knot and link invariants, making a
special emphasis in its latest developments related to string theory.

\section{Topological quantum field theory in two and four dimensions}

Topological quantum field theories are quantum field theories whose
correlation functions lead to topological invariants. They can be
constructed in two ways, leading to two types of theories: Schwarz and Witten
types. In the first case one considers theories whose action is manifestly
independent of the metric. Then, if the observables do not involve the metric
either, correlators lead to topological invariants (if there are no metric
dependences induced upon quantization). An example of Schwarz-type theory is
Chern-Simons gauge theory.

In Witten-type theories, also known as theories of cohomological type,
though the action manifestly depends on the metric, a symmetry prevents
the correlators of being metric dependent. These theories posses a
supersymmetric counterpart and their characteristic symmetry is indeed
related to supersymmetry. Topological quantum field theories of this type
are twisted versions of ordinary ${\cal N}=2$ supersymmetric theories. Under
the twist the Lorentz transformations of some of the fields are modified
and, though on flat space both theories are the same, on curved space are
different. On curved space only one of the supersymmetry transformations
survives as a symmetry. This symmetry is a nilpotent scalar quantity $Q$.
The operators whose correlators lead to topological invariants, or
``physical observables", are the elements of the cohomology generated by
$Q$. In these theories the energy-momentum tensor is $Q$-exact and thus the
elements of the cohomology of $Q$ lead to quantities which are metric
independent (modulo potential anomalies). In these theories the functional
integral is typically localized on field configurations  which are
$Q$-invariant, leading to a moduli problem. The two classical examples of
these types of topological quantum field theories are topological sigma
models in two dimensions \cite{tsm} and topological Yang-Mills theory in four
dimensions \cite{tqft}.

Topological sigma models are obtained after twisting ${\cal N}=2$
supersymmetric sigma models. The twisting can be done in two different ways
leading to two types of models, A and B \cite{aandb}. Their existence is
related to mirror symmetry. Only type-A models will be described in what
follows. These models can be defined on an arbitrary almost complex
manifold, though typically they are considered on  K\"ahler manifolds. The
theory involves maps from two-dimensional Riemann surfaces $\Sigma$ to
target spaces $X$, together with fermionic degrees of freedom on $\Sigma$
which are mapped to tangent vectors on $X$. The functional integral of the
resulting theory is localized on holomorphic maps, defining the
corresponding moduli space. The
$Q$-cohomology provides the set of physical observables. They can be
integrated as classes over the moduli space leading to topological
invariants.

  Topological sigma models keep fixed the complex structure of the Riemann
manifold $\Sigma$. Motivated by string theory one also considers the
situation in which one integrates over complex structures. In this case one
ends working with holomorphic maps in the entire moduli space of curves. The
resulting theories are called topological strings.

I will review now a particular example of topological string theory
which, besides being very interesting from the point of view of physics and
mathematics, it will be very useful in the discussion presented in the next
section. Let us consider topological strings choosing $X$ to be a
Calabi-Yau threefold. In this case the virtual dimension of the moduli space
of holomorphic maps turns out to be zero. Two situations can occur: the
space is given by a number of points or the space involves a moduli and
possesses a bundle of the same dimension as the tangent bundle. In the first
case, topological strings count the number of points weighted by the
exponential of the area of the holomorphic map (the pull back of the
 K\"ahler form integrated over the surface) times $x^{2g-2}$ where $x$ is the
string coupling constant and $g$ is the genus of $\Sigma$. In the second case
one computes the top Chern class of the appropriate bundles (properly
defined), again weighted by the same factor. In both cases one can classify
the contributions according to the cohomology class $\beta$ on $X$ in which
the image of the holomorphic map is contained. The sum of the numbers
obtained for each $\beta$ and fixed $g$ are known as Gromov-Witten
invariants, $N_g^\beta$. The topological string contribution takes the form:
\begin{equation}
\sum_{g\geq 0} x^{2g-2} \Bigg( \sum_{\beta\in H_2(X,Z)}  N_g^\beta
\ex^{\int_\beta
\omega} \Bigg)
\label{gw}
\end{equation}
where $\omega$ is the  K\"ahler class of the Calabi-Yau manifold. In general,
the numbers $N_g^\beta$ are rational numbers.

The discussion has shown how Gromov-Witten invariants can be interpreted in
terms of string theory. One could think that this is just a nice observation
and that no new insight on these invariants could be obtained from this
formulation. The situation turns out to be quite the opposite. Once a string
formulation has been obtained the whole machinery of string theory is at our
disposal. One should look to new ways to compute the quantity (\ref{gw}),
where Gromov-Witten invariants are packed. The hope is that, if this is
possible, the new emerging picture will provide new insights on these
invariants. This is indeed what occurred in the last three years. It turns out
that the quantity (\ref{gw}) can be obtained from an alternative point of
view in which the embedded Riemann surfaces are regarded as D-branes
\cite{gopa}. The outcome of this approach is that the Gromov-Witten
invariants can be written in terms of other invariants which are integers and
that posses a geometrical interpretation. To be more specific, the quantity
(\ref{gw}) takes the form:
\begin{equation}
\sum_{g\geq 0 \atop \beta\in H_2(X,Z)} \sum_{d>0} 
n_g^\beta {1\over d} (2\sin({dx\over 2}))^{2g-2}
\ex^{d\int_\beta \omega}
\label{gwgv}
\end{equation}
where $n_g^\beta$ are the new {\it integer} invariants. This prediction has been
verified in all the cases in which it has been tested \cite{tests}. A
similar structure will be found in the next section in the context of
knot theory in the large-$N$ limit.

\vskip1cm

I will now briefly review topological Yang-Mills theory in four
dimensions. As the topological sigma models, this theory is constructed
by twisting ${\cal N}=2$ supersymmetric Yang-Mills theory. This process
modifies the spin content of the fields of the theory, leading to a new
nonequivalent theory on curved manifolds. Again, out of the full ${\cal
N}=2$ supersymmetry transformations, a nilpotent scalar symmetry is
preserved that guaranties the topological character of the theory. The
corresponding cohomology fixes the physical observables. It turns out that
in this theory the contributions from the functional integral are localized
on the moduli space of instantons. The correlators turn out to be integrals
of appropriate forms on this moduli space, leading to quantities that are
identified with Donaldson invariants.

To be more specific, let us consider the case in which the gauge group
is $SU(2)$, and the four manifold
$X$ is simply connected and has $b_2^+>1$ (the case $b_2^+=1$ is
anomalous). In this situation there are
$1+b_2$ physical observables, ${\cal O}$ and $I(\Sigma_a)$,
$a=1,\dots,b_2$, where ${\Sigma_a}$ is a basis of $H_2(X)$. After packing
these observables in a generating functional, topological Yang-Mills
theory leads to the computation of the following functional integral:
\begin{equation}
\Big\langle \exp \Big(\sum_a \alpha_a I(\Sigma_a) + \lambda {\cal O}
\Big) \Big\rangle,
\label{donstr}
\end{equation}
where $\lambda$ and $\alpha_a$, $a=1,\dots,b_2$, are parameters. In
computing this quantity one can argue that the contribution is localized
on the moduli space of instantons configurations and one ends, after
taking into account the selection rule dictated by the dimensionality of
the moduli space, with integrations over the moduli space of the selected
forms. The resulting quantities are precisely Donaldson invariants.

As in the case of topological sigma models one could be tempted to argue
that the observation leading to a field-theoretical interpretation of
Donaldson invariants does not provide any new insight. It would be a
 mistake to do so. Once a field theory
formulation is available one has at his disposal a huge machinery which
could lead, on the one hand, to further generalizations of the theory and,
on the other hand, to new ways to compute quantities like (\ref{donstr}),
obtaining new insights on these invariants. This is indeed what happened in
the nineties, leading to an important breakthrough in 1994 when Seiberg
and Witten calculated  (\ref{donstr}) in a different way and pointed out
the relation of Donaldson invariants to new integer invariants that
nowadays carry their names.

The localization argument that led to the interpretation of
(\ref{donstr}) as Donaldson invariants is valid because the theory under
consideration is exact in the weak coupling limit. Actually, one can
easily argue that the topological theory under consideration is
independent of the coupling constant and thus calculations in the strong
coupling limit are also exact. These type of calculations were out of the
question before 1994. The situation changed dramatically after the work
by Seiberg and Witten in which ${\cal N}=2$ super Yang-Mills theory was
solved in the strong coupling limit \cite{swpaper}. Its application to  the
corresponding twisted version was immediate and it turned out that
Donaldson invariants can be written in terms of new integer invariants
now known as Seiberg-Witten invariants \cite{mono}. The development has a
strong resemblance with the one described above for topological strings:
certain non integer invariants can be expressed in terms of new integer
invariants.

The Seiberg-Witten invariants are actually simpler to compute than
Donaldson invariants. They correspond to partition functions of
topological Yang-Mills theories where the gauge group is abelian. These
contributions can be grouped into classes labeled by $x=-2c_1(L)$ where
$c_1(L)$ is the first Chern class of the corresponding line bundle. The
sum of contributions, each being $\pm 1$, for a given class $x$ is the
{\it integer} Seiberg-Witten invariant $n_x$. The strong coupling analysis of
topological Yang-Mills theory leads to the following expression for
(\ref{donstr}):
\begin{equation}
2^{\scriptstyle{1+{1\over4}(7\chi+11\sigma)}}
\left(\ex^{\left({{v^2}\over2}+2\lambda\right)}
\sum_x n_x\ex^{\scriptstyle{v\cdot x}}+i^{\chi+\sigma\over 4}  
\ex^{\left(-{v^2\over2}-2\lambda\right)}
\sum_x n_x\ex^{\scriptstyle{-iv\cdot x}}\right).\nonumber\\
\label{sw}
\end{equation}
where $v=\sum_a \alpha_a \Sigma_a$, and $\chi$ and $\sigma$ are the Euler
number and the signature of the manifold $X$. This result matches the known
structure of (\ref{donstr}) (structure theorem of Kronheimer and Mrowka
\cite{kron}) and provides a meaning to its unknown quantities in terms of
the new Seiberg-Witten invariants. Equation (\ref{sw}) is a rather
remarkable prediction that has been tested in many cases. A general proof
of the relation between Donaldson and Witten invariants has been proposed
recently in \cite{feh}. For a review of the subject see \cite{lectu}.

The situation for manifolds with $b_2^+=1$ involves a metric dependence
and has been worked out in detail in \cite{moorewitten}.
The formulation of Donaldson invariants in field-theoretical terms has
also provided a generalization  of these invariants. This generalization
has been carried out in several directions: {\it a)} the consideration of
higher-rank groups \cite{marmoore}, {\it b)} the coupling to matter
fields after twisting ${\cal N}=2$ hypermultiplets \cite{labmar}, {\it c)}
the twist of theories involving ${\cal N}=4$ supersymmetry for gauge group
$SU(2)$
\cite{vawi}, and for higher rank groups \cite{loza}.

\section{Chern-Simons gauge theory}

In this section I will briefly review the most significant topological
quantum field theory in three dimensions. 
Chern-Simons gauge theory  is a
topological  quantum field theory whose action is built out of a
Chern-Simons term involving as gauge field a gauge connection associated
to a group $G$ on a three-manifold $M$,
\begin{equation}
 S={{k}\over 4\pi} \int_M {\rm Tr} \Bigl( A \wedge
d A + {2
\over 3} A
\wedge A \wedge A \Bigr),
\label{csaction}
\end{equation}
where $k$ is an integer. The natural physical observables of the theory
are Wilson loops:
\begin{equation}
W^{K}_{R}(A)={\rm Tr}_{R} {\rm P}\,\exp\,
\oint_{K} A,
\label{wilson}
\end{equation}
where $K$ is a loop and $R$ a representation of the gauge group. The
vacuum expectation values of products of these operators are topological
invariants which are related to quantum-group invariants. Given a link
${\cal L}$ of $L$ components, $K_1,K_2,\dots,K_L$, one computes
correlators of the form $\langle  W^{{K}_1}_{
R_1}\cdots W^{{K}_L}_{ R_L}\rangle $, where $R_1,R_2,\dots,R_L$ are
representations associated to each component. These quantities turn out
to be polynomials in $q=\ex^{2\pi i\over k+N}$ and $\lambda=q^N$ with
integer coefficients. In what follows I will concentrate on the case of
knots, denoting the vacuum expectation value of the Wilson loop
(\ref{wilson}) simply by $W_R$.

The observation that these correlators lead to knot and link
invariants was done by Witten in 1988 \cite{cs}. He provided a completely
new point of view for theses invariants which turned out to be very
fruitful over the years. The first study of this theory used
non-perturbative field theoretical methods and led to the identification
of its correlators with quantum-group invariants for knots and links
\cite{qg}. Again, as in the two cases discussed in the previous section,
the different methods of field theory were available to study the
theory from different perspectives. As it will be now briefly reviewed, the
progress made in this case is even more impressive that in the previous
cases. 

Besides non-perturbative methods,
perturbative ones can be also applied. These were soon developed for
Chern-Simons gauge theory \cite{pert}, and they provided important
representations of Vassiliev invariants. These invariants, proposed by
Vassiliev in 1990 \cite{vass}, turned out to be the coefficients of the
perturbative series expansion of the correlators of Chern-Simons gauge
theory \cite{birman}. Perturbative studies can be carried out in different
gauges, originating a variety of new representations of Vassiliev
invariants. Among the more relevant results related to these topics are
the integral expressions for Vassiliev invariants by Kontsevich
\cite{kont} and by Bott and Taubes \cite{bt}, as well as the recent
combinatorial ones \cite{combi} in the spirit of \cite{arrgpo}. I will
not describe these developments here but refer the interested reader
to the recent review
\cite{laplata}. In this paper I will concentrate in the new perspective
emerged after studying the large
$N$ expansion of the theory. I will restrict the discussion to the case
of knots on $S^3$ with gauge group $SU(N)$.

Gauge theories with gauge group $SU(N)$ admit, besides the perturbative
expansion, a large-$N$ expansion. In this expansion correlators are
expanded in powers of $1/N$ while keeping the 't Hooft coupling $t=N x$
fixed, being
$x$  the coupling constant of the gauge theory. For example, for the
free energy of the theory one has the general form,
\begin{equation}
F=\sum_{g\ge 0 \atop h\ge 1}^\infty C_{g,h} N^{2-2g} t^{2g-2+h}.
\label{largen}
\end{equation}
In the case of Chern-Simons gauge theory, the coupling constant is
$x={2\pi i \over k+N}$ after taking into account the shift in $k$
\cite{shift}. The large-$N$ expansion (\ref{largen}) resembles a string
theory expansion and indeed the quantities $C_{g,h}$ can be identified
with the partition function of a topological open string with $g$ handles
and $h$ boundaries, with $N$ D-branes on $S^3$ in an ambient
six-dimensional target space $T^*S^3$. This was pointed out by Witten in
1992
\cite{witdb}. The result makes a connection between a topological
three-dimensional field theory and the topological strings described in
the previous section.

In 1998 an important breakthrough took place which provided a new
approach to compute quantities like  (\ref{largen}). Using arguments
inspired by the AdS/CFT correspondence (see
\cite{ads} for a review), Gopakumar and Vafa
\cite{gova} provided a closed string theory interpretation of the
partition function (\ref{largen}). They conjecture that the free energy
$F$ can be expressed as,
\begin{equation}
F=\sum_{g\ge 0 }^\infty  N^{2-2g} F_g(t),
\label{closed}
\end{equation}
where $F_g(t)$ correspond to the partition function of a topological
closed string theory on the non-compact Calabi-Yau manifold $X$ called
the resolved conifold,
${\cal O}(-1) \oplus {\cal O}(-1) \rightarrow {\bf P}^1,$
being $t$ the flux of the $B$-field through ${\bf P}^1$.
The quantities $F_g(t)$ have been computed using physical \cite{gova} and 
mathematical arguments \cite{rig}, proving the conjecture.

Once a new picture for the partition function of Chern-Simons gauge theory
is available one should ask about the form that the expectation values of
Wilson loops could take in the new context. The question was faced by
Ooguri and Vafa and they provided the answer \cite{oova}, later refined in
\cite{lmv}. The outcome is an entirely new point of view in the theory of
knot and link invariants. The new picture provides a geometrical
interpretation of the integer coefficients of the quantum group invariants,
an issue that has been investigated during many years. To present an account
of these developments one needs to review first some basic facts of
large-$N$ expansions.

To consider the presence of Wilson loops it is convenient to introduce a
particular generating functional. First, one performs a change
basis from representations $R$ to conjugacy classes $C(\vec k)$ of the
symmetric group, labeled by vectors $\vec k= (k_1,k_2,\dots)$ with $k_i\ge
0$, and
$|\vec k| = \sum_j k_j > 0$. The change of basis is $W_{\vec k} = \sum_R
\chi_R(C(\vec k)) W_R$, where $\chi_R$ are characters of the permutation
group $S_\ell$ of
$\ell=\sum_j j k_j$ elements ($\ell$ is also the number of boxes of the
Young tableau associated to $R$). Second, one introduces
the generating functional:
\begin{equation}
F(  V )=\log Z(  V  )= \sum_{\vec k}
{|C({\vec k} )|
\over   \ell  !}  W^{  (c)}_{{\vec k} } 
\Upsilon_{{\vec k}}(  V ),
\label{genfun}
\end{equation}
where $Z(  V  )= \sum_{\vec k}
{|C({\vec k} )|
\over   \ell  !}  W^{}_{{\vec k} } 
\Upsilon_{{\vec k}}(  V )$ and $\Upsilon_{\vec{k}}(
V)=\prod_j ({\rm Tr} V^{ j})^{k_{ j}} $. In these expressions $|C({\vec
k} )|$ denotes the number of elements of the class $C({\vec k} )$ in
$S_\ell$. The reason behind the introduction of this generating
functional is that the large-$N$ structure of the connected Wilson
loops, $W^{  (c)}_{\vec{k} }$, turns out to be very simple:
\begin{equation}
{  |C(\vec{k} )| \over  \ell  !}
W^{  (c)}_{\vec{k} } =\sum_{g=0}^{\infty}  x^{  2g-2+
|\vec{k} |} 
F_{g, \vec{k} }(\lambda ),
\label{largenwil}
\end{equation}
where $\lambda = \ex^t$ and $t=N x$ is the 't Hooft coupling. Writing
$x=t/N$, it corresponds to a power series expansion in $1/N$. As before,
the expansion looks like a perturbative series in string theory where $g$
is the genus and $|\vec k|$ is the number of holes. Ooguri and Vafa
conjectured in 1999 the appropriate string theory description of
(\ref{largenwil}). It correspond to an open topological string theory
(notice that the ones described in the previous section were closed) whose
target space is the resolved conifold $X$. The contribution from this
theory will lead to open-string analogs of Gromov-Witten invariants.

In order to describe in more detail the fact that one is dealing with open
strings, some new data needs to be introduced. Here is where the knot
intrinsic to the Wilson loop enters. Given a knot $K$ on $S^3$, let us
associate to it a Lagrangian submanifold $C_K$ with $b_1=1$ in the resolved
conifold $X$ and consider a topological open string on it. The
contributions in this open topological string are localized on holomorphic
maps $f:\Sigma_{g,h}
\rightarrow X$ with $h=|\vec k|$ which satisfy: $f_*[\Sigma_{g,h}]={\cal
Q}$, and $f_*[C]=j[\gamma]$ for $k_j$ oriented circles $C$. In these
expressions $\gamma\in H_1(C_K,{\bf Z})$, and ${\cal Q}\in H_2(X,C_K,{\bf
Z})$, \ie, the map is such that $k_j$ boundaries of $\Sigma_{g,h}$ wrap
the knot $j$ times, and $\Sigma_{g,h}$ itself gets mapped to a relative
two-homology class characterized by the Lagrangian submanifold $C_K$.
The number of these maps (in the sense described in the previous section)
constitute the open-string analogs of Gromov-Witten invariants. They will
be denoted by $N_{g,\vec k}^{\cal Q}$. Comparing to the situation that
led to (\ref{gw}) in the closed string case one concludes that in this
case the quantities $F_{g, \vec{k} }(\lambda )$ in (\ref{largenwil}) must
take the form:
\begin{equation}
F_{g, \vec{k} }(\lambda ) = \sum_{\cal Q} N_{g,\vec k}^{\cal Q}
\ex^{\int_{\cal Q} \omega}, \,\,\,\,\,\,\,\,\,\,\,\,\,\,\,\,
t=\int_{{\bf P}^1} \omega,
\label{gwopen}
\end{equation}
where $\omega$ is the  K\"ahler class of the Calabi-Yau manifold $X$ and
$\lambda = \ex^t$. For any ${\cal Q}$, one can always write $\int_{\cal
Q} \omega = Qt$ where $Q$ is in general a half-integer number. Therefore,
$F_{g, \vec{k} }(\lambda )$ is a polynomial in $\lambda^{\pm {1\over 2}}$
with rational coefficients.

The result (\ref{gwopen}) is very impressive but still does not provide a
representation where one can assign a geometrical interpretation to the
integer coefficients of the quantum-group invariants.
Notice that to match a polynomial invariant to (\ref{gwopen}), after
obtaining its connected part, one must expand  it in $x$ after setting
$q=\ex^x$ keeping $\lambda$ fixed. One would like to have a refined version
of (\ref{gwopen}), in the spirit of what was described in the previous
section leading from the Gromov-Witten invariants $N_g^\beta$ of
(\ref{gw}) to the new integer invariants $n_g^\beta$ of (\ref{gwgv}).
This study was indeed done in \cite{oova} and later
improved in \cite{lmv}. The outcome is that, indeed, $F(V)$ can be
expressed in terms of integer invariants in complete analogy with the
description presented in the previous section for topological strings.

In order to present these results one needs to perform first a
reformulation of the quantum-group invariants or vacuum expectation
values of Wilson loops. Instead of considering $W_R(q,\lambda)$, a
corrected version of it, $f_R(q,\lambda)$, will be studied. These
reformulated polynomial invariants have the form:
\begin{eqnarray}
 f_{ R }(q, \lambda) &=&
 \sum_{d, m=1}^{\infty} (-1)^{m-1} {\mu 
(d) \over d m} 
\sum_{ \{ \vec{k}^{(j)},  R_{ j} \} } 
 \chi_{ R} \biggl( 
C\biggl( (\sum_{j=1}^m \vec{k}^{(j)})_d\biggr)\biggr)\nonumber\\
&& \times \prod_{j=1}^m {|C(\vec{k}^{(j)})| \over  \ell_{
 j}!}
  \chi_{ R_{ j}}(C(\vec{k}^{(j)})) 
W_{ R_{ j}}(q^{ d}, \lambda^{ d}),
\label{lafor}
\end{eqnarray}
where $({\vec k}_d)_{di} = k_i$ and zero otherwise. In this expression
$\mu(d)$ is the Moebius function. In spite of its frightening form, the
reformulated polynomial invariants $f_R(q,\lambda)$ are just
quantum-group invariants $W_R(q,\lambda)$ plus lower order terms,
understanding by this terms which contain quantum-group invariants
carrying representations whose associated Young tableaux have a lower
number of boxes. For example, for the simplest cases:
\begin{eqnarray}
f_{\tableau{1}}(q,\lambda)&=&W_{\tableau{1}}(q, \lambda), \nonumber\\
f_{\tableau{2}}(q,\lambda)&=&W_{\tableau{2}}(q,\lambda)
-{1\over 2}\bigl( W_{\tableau{1}}(q,\lambda)^2+ 
W_{\tableau{1}}(q^2,\lambda^2)
\bigr),\nonumber\\
f_{\tableau{1 1}}(q, \lambda)&=&W_{\tableau{1 1}}(q,\lambda)
-{1\over 2}\bigl( W_{\tableau{1}}(q,\lambda)^2-
 W_{\tableau{1}}(q^2,\lambda^2) 
\bigr).
\label{examples}
\end{eqnarray}
The nice feature of the reformulated quantities is that $F(V)$
acquires a very simple form in terms of them:
\begin{equation}
F(V)=\sum_{d=1}^\infty \sum_R {1\over d} f_R(q^d,\lambda^d) {\rm Tr} V^d,
\label{simple}
\end{equation}
and therefore they seem to be the right quantities to express an
alternative point of view in which the embedded Riemann surfaces can be
regarded as D-branes (recall equation (\ref{gwgv})).

To present the conjectured form of the reformulated $f_R(q,\lambda)$ a
couple of new ingredients are needed. One needs the Clebsch-Gordon
coefficients $C_{R,R',R''}$ of the symmetric group (they satisfy
$V_R\otimes V_{R'} = \sum_{R''} C_{R,R',R''} V_{R''}$), and  monomials
$S_R(q)$ defined as follows: $S_R(q)=(-1)^d q^{d-{\ell-1\over 2}}$ if $R$
is a hook representation, with $\ell-d$ boxes in the first row, and
$S_R(q)=0$ otherwise. The conjecture presented in
\cite{lmv} states that the reformulated invariants have the form:
\begin{equation}
f_{  R}  (q ,
\lambda )=\sum_{g\ge 0}\sum_{Q,  R' ,   R''} 
C_{R R' R''}  N_{  R' , g, Q} S_{ 
R'' }(q )(q^{ {1
\over 2}} - q ^{  -{1
\over 2}})^{2g-1} \lambda^{  Q},
\label{conjecture}
\end{equation}
where $N_{  R , g, Q}$ are {\it integer} invariants which posses a geometric
interpretation. These quantities are the analog of the integers
$n_g^\beta$ in Gromov-Witten theory. They can be described in terms of
the moduli space of Riemann surfaces with boundaries embedded into a
Calabi-Yau manifold. Their geometrical interpretation has been treated
recently in \cite{last}. 

The structure (\ref{conjecture}) has been verified for a variety of
non-trivial knots and links, and representations up to four boxes
\cite{testing,lmv}. For the unknot, the whole picture have been verified in
complete detail \cite{oova,nonudo}. The form of $F(V)$ for the unknot can be
easily computed in Chern-Simons gauge theory,
\begin{equation}
  F(  V )=\sum_{d=1}^{\infty} { \lambda^{d  \over2} - 
\lambda^{-{d \over 2}} 
\over 2d \sin \bigl( {d x   \over 2} \bigr)} {\rm
Tr}_{ \tableau{1}} {  V }^d,
\label{unknot}
\end{equation}
leading to an expansion (\ref{largenwil}) of the form:
\begin{equation}
F_{g, (  0  ,\cdots,   0 ,  1 ,
  0 ,
\cdots,
  0 )}(\lambda)={ (1-2^{1-2g})|B_{2g}|
\over  (2g)! }d^{2g-2} (\lambda^{d  \over2} - 
\lambda^{-{d \over 2}}), 
\label{bern}
\end{equation}
where  the 1 in $F_{g, (  0  ,\cdots,   0 ,  1 ,
  0 ,
\cdots,
  0 )}$ is located in the $d^{\rm th}$ position. Form these equations one can
easily read the numbers which correspond to the open-string analogs of
Gromov-Witten invariants,
$N_{g,\vec k}^{\cal Q}$, in (\ref{gwopen}), as well as the new integer
invariants present in the general expression (\ref{conjecture}):
\begin{equation}
N_{\tableau{1},0,{1\over 2}} = - N_{\tableau{1},0,-{1\over 2}}=1.
\label{prims}
\end{equation}

In this section I have shown how the string theory description of the
large-$N$ expansion of Chern-Simons gauge theory provides a 
new point of view in the study of knot and link invariants.
Quantum-group polynomial invariants are
reformulated so that their integer coefficients and exponents possess a
geometric interpretation.
 The new integer invariants are identified in terms
of topological properties of the moduli space of
Riemann surfaces with boundaries embedded into a Calabi-Yau manifold.
The new framework provides strong predictions on
the  algebraic structure of the quantum group
polynomial invariants.

Many open problems remain to be studied. One would
like to know what are the implications of the skein rules on the new
invariants, and, on the contrary, one would like to understand the new algebraic
structure of the polynomial invariants from the point of view of quantum
groups.  Also, one should
study the extension of the formalism to other gauge groups. From a more
mathematical perspective, a detailed analysis of the moduli spaces
involved  in the computation of the new invariants is needed, as well as
the development of computational technics.  One should also analyze the
picture emerging from the theories which are mirror to the topological
strings involved. Some work in this direction has been done recently
in \cite{mirr}.

\vskip0.6cm

 In this paper I have described how the many faces of quantum field theory
and string theory  provide a variety of important insights in a selected
set of problems in topology. It is particularly remarkable the connection
between the first and the third contexts considered. One observes a
fascinating interplay between string theory, knot theory and enumerative
geometry which opens new fields of study.

\vskip 0.3in 
\vbox{\centerline{\bf Acknowledgments}
\bigskip

I would like to thank the organizers of the special seminar in honor of
F. J. Yndur\'ain for inviting me to deliver a talk at the meeting held
at Sitges, Spain, on February 9, 2001, in the context of the 
XXIX International Meeting on Fundamental
    Physics. I would like to
thank also M. Alvarez, C. Lozano, P.M. Llatas, E. P\'erez, A. V. Ramallo and,
specially, M. Mari\~no,  for collaborations and discussions on many of the
topics described here. This work is supported in part by Ministerio de Ciencia y
Tecnolog\'\i a under grant PB96-0960, and by Xunta de Galicia under grant
PGIDT00-PXI-20609. }

\vfill
\newpage

\vskip 0.6in 
\vbox{\centerline{\bf Afterword}
\bigskip

This paper is dedicated to Francisco J. Yndur\'ain, Paco, as his friends use
to call him. Paco was my first quantum field theory teacher in the Universidad
Aut\'onoma de Madrid. I learnt from him the basics of the subject, a topic
that he mastered and enjoyed. He is an excellent teacher, emphasizing the
essentials and avoiding cumbersome detours. These features are clear in his
textbooks, pieces of great work already enjoyed by several generations. I
also started with Paco my career as a researcher. I wrote with him my first
paper, dealing with some aspects of proton decay. That paper contains the only
measurable physics prediction in which I has been involved. Unfortunately, it
has been ruled out. I have made, however, testable mathematical
predictions that have been verified (the risk is much lower). As a young
researcher, I learnt from Paco that one should always try to face important
problems no matter how hard or fashionable they are. Paco has certainly made
many valuable contributions to theoretical physics. I am sure that he will
continue doing so for many years. He has also made important efforts to promote
high-energy physics in Spain (both, theoretical and experimental), to rise it
to the level that we enjoy nowadays. In this task, he was always guided by
high-quality standards. We should all thank him for his firm commitment.}

\end{document}